# A Robust and Efficient Node Authentication Protocol for Mobile Ad Hoc Networks

Jaydip Sen
Innovation Labs, Tata Consultancy Services Ltd.
Bengal Intelligent Park, Salt Lake Electronic Complex
Kolkata-700091, INDIA
E-mail: Jaydip.Sen@tcs.com

*Abstract*— A *mobile ad hoc network* (MANET) is a collection of mobile nodes that communicate with each other by forming a multi-hop radio network. Security remains a major challenge for these networks due to their features of open medium, dynamically changing topologies, reliance on cooperative algorithms, absence of centralized monitoring points, and lack of clear lines of defense. Design of an efficient and reliable node authentication protocol for such networks is a particularly challenging task since the nodes are battery-driven and resource constrained. This paper presents a robust and efficient key exchange protocol for nodes authentication in a MANET based on multi-path communication. Simulation results demonstrate that the protocol is effective even in presence of large fraction of malicious nodes in the network. Moreover, it has a minimal computation and communication overhead that makes it ideally suitable for MANETs.

## I. INTRODUCTION

A MANET is a collection of wireless hosts that can be rapidly deployed as a multi-hop packet radio network without the aid of any established infrastructure or centralized administrator. Such networks can be used to enable next generation battlefield applications, including situation-aware systems for maneuvering war fighters, and remotely deployed unmanned micro-sensor networks. MANETs have some special characteristic features such as unreliable wireless media (links) used for communication between hosts, constantly changing network topologies and memberships, limited bandwidth, battery, lifetime, and computation power of nodes etc. While these characteristics are essential for the flexibility of MANETs, they introduce specific security concerns that are absent or less severe in wired networks. MANETs are vulnerable to various types of attacks. These include passive eavesdropping, active interfering, impersonation, and denial-of-service. Intrusion prevention measures such as strong authentication and redundant transmission can be used to improve the security of a MANET. However, these techniques can address only a subset of the threats. Moreover, they are costly to implement. The dynamic nature of MANETs requires that prevention techniques should be complemented by detection techniques, which monitor security status of the network and identify malicious behavior.

One of the most critical problems in MANETs is the security vulnerabilities of routing protocols. A set of nodes in a MANET may be compromised in such a way that it may not be possible to detect their malicious behavior easily. Such nodes can generate new routing messages to advertise non-existent links, provide incorrect link state information, and flood other nodes with routing traffic, thus inflicting Byzantine failure in the network. Several secure routing protocols have been proposed for MANETs based on cryptographic mechanisms [1]. Almost all of them assume the existence of a secure channel through which a security association has been established between the source and the destination. However, the prerequisite for such a secure channel to exist is the existence of a security association. This creates a *routing security interdependency cycle* [2].

In this paper, an efficient key exchange protocol is proposed for MANETs that can be easily integrated with a routing protocol thereby providing an integrated framework of routing and security and solving the routing security interdependency cycle. The main contributions of the paper are as follows: (i) it presents a reliable and efficient key exchange protocol for MANETs, (ii) the protocol is based on multi-path communication, and therefore, it is robust even in presence of malicious nodes in the network, and (iii) the protocol involves minimal computation overhead and is ideal for resource-constrained nodes in MANETs. The rest of this paper is organized as follows. Section II presents some related work in MANET security. Section III describes the proposed protocol. Section IV provides performance evaluations of the protocol through simulations. Finally, Section V concludes the paper while highlighting some future scope of work.

## II. RELATED WORK

The problem of security and cooperation enhancement among the nodes in a MANET has received considerable attention by the researchers over the last decade. Cryptography remains the most widely proposed solution to provide authentication for nodes. However, cryptography assumes safe key-exchange, which is particularly difficult to realize in open multi-hop network communications where attackers may launch man-in-the-middle attacks.

Zhou et al. have introduced a threshold cryptography-based key management scheme for MANETs [3]. A group of

*n* servers together with a master public/private key pair are first deployed by a *certificate authority* (CA). Each server has a share of the master private key and stores the key pairs of all nodes. The shares of the master private key are generated using *threshold cryptography*. Thus only *n* servers together can form a complete signature. If any node wants to join the network, it must first collect all of the *n* partial signatures. Then the node can compute the complete signature locally and get the certificate. This scheme has been extended in a mechanism proposed by Kong et al. [4], where a *centralized dealer* is introduced to issue certificates and private key shares to *t* nodes during the network bootstrapping phase. A threshold cryptography system is deployed in order to provide a ($t$, $n$) secret sharing service. Any t nodes can form a centralized dealer and can issue or revoke certificates. In this scheme, any node that is willing to join the network will have to collect *t* partial signatures in its local communication range. Although, this scheme, to some extent, can handle the issue of nodes' joining and leaving the network, it increases the risk of leaking of the private key of the centralized dealer. In the event of *t* nodes being compromised, the security of the entire network will be in jeopardy.

Montenegro et al. have proposed a *statistically unique and cryptographically verifiable* (SUCV) identifier scheme in which the nodes compute their addresses applying a non-reversible hash function on their public key [5]. Any node can then directly bind a public key to its owner address and an IP can not be spoofed without the associated private key. This provides a reliable authentication scheme for the nodes in a MANET.

Hubaux et al. have proposed a self-organized public key infrastructure-based trust building scheme [6] for MANETs by adapting *pretty good privacy* (PGP) protocol [7]. However, unlike in PGP, in this scheme, there are no central certificate directories for distribution of certificates. In order to find the public key of a remote user, a local user makes use of *Hunter Algorithm* [6] on the merged certificate repository to build certificate chain(s). A certificate chain initiates from the local user's certificate and terminates at the remote user's certificate. The probability of finding such a certificate chain in this scheme is high, but it is not guaranteed. Although the scheme is very suitable for decentralized networks like MANETs, it leads to disclosure of too much information about the originating nodes, as it releases several unnecessary certificates.

Eshenaur et al. have proposed a trust establishment mechanism in which a node in a MANET can generate trust evidence about any other node [8]. In this scheme, when a *principal node* generates a piece of trust evidence, it signs the evidence with its own private key, specifying the lifetime and makes it available to others through the network. A principal may revoke a piece of evidence it produced by generating a revocation certificate for that piece of evidence and making available to others at any time before the evidence expires. A principal can get disconnected after distributing trust evidence. Similarly, a producer of trust evidence does not have to be reachable at the time its evidence is being evaluated. Evidences can be replicated across various nodes to guarantee availability. Although the scheme is conceptually sound, the authors have not provided any details about the performance evaluation of the scheme.

Abdul-Rahman et al. have proposed a *distributed trust model* - a decentralized approach to trust management that uses a recommendation protocol to exchange trust-related information [9]. The model assumes that relationships are unidirectional and exist between and two entities (nodes). The entities make judgments about the quality of a recommendation of trust, based on their policies, i.e., they have values for trust relationships. The recommendation protocol works by requesting a trust value in a trust target with respect to a particular classification. After getting an answer, an evaluation function is used to obtain an overall trust value in the target node. The protocol also allows recommendation refreshing and revocation, and is suited for establishing trust relationships that are less formal and temporary in nature.

Asokan et al. have introduced several password-based key exchange methods to set up a secure session among a group of nodes without any support infrastructure [10]. In this scheme, only those nodes that know an initial password are able to obtain the session key. The session key is formed by combinations from all the nodes in the network. The basic working principle of the scheme is as follows. A weak password is sent to the group members. Each member then contributes to generation of part of the key and signs this data by using the weak password. Finally, to establish a secure session key, a secure channel is derived without any central trust authority or support infrastructure.

Stajano et al. have introduced the *resurrecting duckling* security protocol to establish trust in ad hoc networks [11]. The protocol is particularly suited for devices without display and for embedded devices that are too weak for public key operations. The fundamental authentication problem is solved by a secure transient association between two devices establishing a master-slave relationship. It is secure in the sense that the master and the slave share a common secret. The protocol is transient because the master at any point of time can terminate the association.

Repantis et al. have proposed a decentralized trust management middleware for ad hoc, peer-to-peer networks based on *reputation* of nodes [12]. The reputation information of each peer is stored in its neighborhood and piggybacked on its replies.

Patwardhan et al. have proposed a trust-based data management scheme in which mobile nodes access distributed information, storage and sensory resources available in pervasive computing environment [13]. The authors have taken a holistic approach that considers data, trust, security, and privacy and utilizes a collaborative mechanism that provides trustworthy data management platform in an ad hoc network.

Baras et al. have proposed a trust management scheme for self-organized ad hoc networks, where the nodes share trust information only with their neighbors [14]. For establishing and maintaining trust among the neighbors, the authors have proposed a voting mechanism. This voting mechanism has made the scheme robust.

Chang et al. have proposed a trust-based scheme for multicast communication in a MANET [15]. In a multicast MANET, a sender node sends packets to several receiving nodes in a multicast session. Since the membership in a multicast group changes frequently in a MANET, the issues of supporting secure authentication and authorization in a multicast MANET is very critical. The proposed scheme involves a two-step secure authentication method. First, an ergodic continuous Markov chain is used to determine the trust value of each one-hop neighbor. Second, the node with the highest trust value is selected as the certificate authority (CA) server. The analytical trust value of each mobile node is found to be very close to that observed in the simulation under various scenarios. The speed of the convergence of the analytical trust value shows that the analytical results are independent of the initial values and the trust classes.

Sun et al. have presented trust as a measure of uncertainty [16]. Using the theory of entropy, the authors have developed techniques to compute the trust values from certain observations. In addition, trust models- entropy-based and probability-based, are presented to solve the concatenation and multi-path trust propagation problems.

Sen et al. have proposed a self-organized trust establishment scheme for nodes in a large-scale MANET in which a trust initiator is introduced during the network bootstrapping phase [17]. The authors have also proposed a distributed trust-based intrusion detection system for MANETs based on cooperation among nodes [18].

However, most of the above-mentioned key management protocols have limitations. In the distributed CA scheme involving threshold cryptography, the trust between a new node willing to join the network and t existing nodes in the network can be established by out-of-bound physical proofs, such as, human perception or biometrics. However, these methods are not very practical in real-world scenario. It may be very difficult, if not impossible, for a node to acquire t existing nodes in the network in its neighborhood for evaluation of its trust. Alternatively, there must be off-line trust establishment mechanism between the new node and t existing nodes. In an infrastructure-less ad hoc network environment, this may also be very difficult to realize in practice. Moreover, threshold cryptography-based schemes have high communication and computation overhead [19].

In self-organized schemes, trust is established through off-line trust relationships among the nodes. These off-line trust relationships are generated from general social relationships. The initialization process depends on the issues of certificates among the nodes themselves and formation of a network of trust relationship between them. This process is very complex and slow in practice, because every issued certificate will require close contact between a pair of nodes. Moreover, ad hoc networks are formed at random by member nodes, and the trust relationships among the members are much sparse than those of general society.

### III. PROPOSED KEY EXCHANGE PROTOCOL

The proposed protocol integrates a key exchange protocol with routing in a MANET and thus solves the *routing-security interdependency* cycle [2]. The objective of a routing protocol is to establish a path between a source node and a destination node. To achieve this objective, the reactive routing protocols for MANETs broadcast a *route request* message in the network so that the route to the destination may be discovered. The proposed key exchange protocol utilizes this approach to retrieve the public keys of the nodes. To find a certificate of a public key, the source node floods the network with a *certificate request* that is replied either by the target node or by an intermediate node that has a valid certificate of the public key of the target node. The proposed protocol is secure against malicious attackers that may try to distribute spurious certificates in the network and cause routing disruption. To make the protocol robust and reliable, two approaches are taken: (i) multi-path certificate exchange and (ii) trust-based certification. The details of the algorithm are described below.

#### A. Description

In the proposed protocol, it is assumed that every node in a MANET first generates a public/private key pair. Since this key pair is generated by the node itself, the node needs to authenticate with some members in the network before it can join and access network resources. This authentication is based on a certificate exchange. The authentication is mutual. Thus, if a node *S* receives the public key of a node *D*, *S* issues a certificate for *D*'s public key. In turn, *D* also issues a certificate for *S*'s public key. In the rest of the paper, the set of nodes that has certified for node *S*'s public key is denoted as *K(S)*. As the authentication is mutual, every node in *K(S)* has its public key certified by *S*.

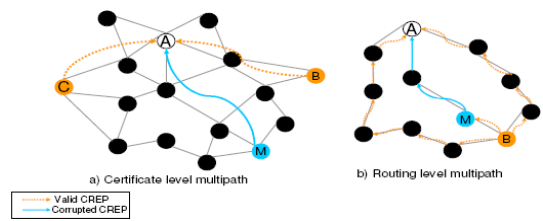

Figure 1. Certificate level multi-path and routing level multi-path

Although the approach of multi-path has not been widely used in certificate exchange schemes for MANETs, it can greatly improve the reliability of a certificate exchange protocol. In designing the proposed protocol two types of multi-path message exchanges are distinguished: (i) multi-path certificate exchange and (ii) multi-path routing. In multi-path certificate exchange, the public key of a node is certified by different nodes (Figure 1 (a)). Due to multiple independent certifications, the confidence assigned to these certificates is higher. A formal computation of the trust values for the certificates may be done using *Dempster-Shafer theory* [6]. Figure 1 (b) shows an example of a multi-path routing, where a node sends a certificate for another node through multiple node-disjoint paths. Since paths do not have any common node, a malicious node can at most prevent a certificate exchange but cannot spoof the identity of another node during the certificate exchange process.

The proposed protocol also utilizes a trust management mechanism [20] to keep track of certification operations. The trust value of a certificate issuing node increases as more number of nodes confirm the public key for which the certificate is issued. On the other hand, when it is detected that a node has issued a spurious certificate, the trust assigned to the node will be decreased and all subsequent certificates issued by the node will also have less confidence associated with them. Consequently, if there is a conflict between certificates, the public key certified by the more trustworthy node(s) will be accepted as genuine.

### B. Operations

*1. Initialization*: In the proposed protocol, before a node enters the network, it generates a public/private key pair. As a node joining for the first time attempts to get several certificates for of its PK from the existing nodes, it floods the network with a *certificate request* (CREQ) message.

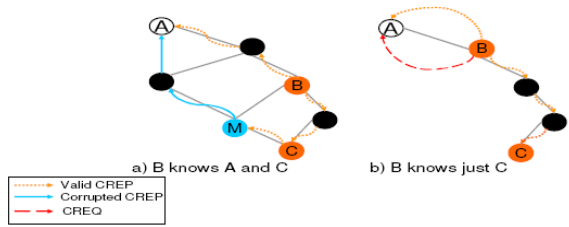

Figure 2. The use of intermediate nodes for certification

*2. Certificate exchange*: Before requesting the node *D*'s certificate, the node *S* evaluates the minimum trust value that is required to consider the public key of *D* as reliable. This threshold value of trust is called the *minimum public key trust value* (MPKTV). This evaluation is local and based on *S*'s security requirements. Node *S* then broadcasts a CREQ for *D*'s certificates including *D*'s address, the list of nodes $K(S)$. The CREQ is sent with a small *time to live* (TTL) to reduce communication overhead of the protocol. Every intermediate node *I* that receives the CREQ checks the public key of *D* and searches its own certificate list.

If *I* has no certificate for *D*, or if it already replied to the CREQ, it simply forwards the packet. Otherwise, *I* sends a *certificate reply* (CREP) to *S* containing a certificate of *D*'s public key signed by *I* (Figure 2(a)). If *I* does not know *S*, it constructs a self-signed certificate and informs *S* that it wants to make a certificate exchange (Figure 2(b)). This packet is sent through multiple node-disjoint paths to *S*. If *I* has a route to *D* in its cache, it informs *D* that *S* has requested its public key. *D* responds and requests a certificate for *S*'s public key. Since *I* and *D* can authenticate each other, the communication between *D* and *I* can be made secure by using *I*'s signature. Therefore, no node can corrupt the certificate of *S* issued by *I*. If *D* does not know sufficient number of nodes, it replies to the CREQ itself.

*S* repeats the operation with an increased TTL until it receives the required minimum number of certificates for *D*'s public key. After receiving the certificates, *S* sends the first packet to *D* which includes the list of nodes which has provided the certificates for *D*'s public key. In this way, *D* gets the information about the known certifiers of *S*. Once they have exchanged their public keys, *S* and *D* issue certificates for each other. This certificate exchange protocol can now be directly applied in routing as *S* and *D* do not have to execute any expensive route discovery procedure for routing.

*3. Certificate revocation*: As authentication is mutual, nodes maintain a list of certifiers. An implicit revocation scheme [6] is adopted, where each node periodically updates its public key by communicating secure certificate exchange messages with its peers.

## IV. PERFORMANCE EVALUATION

### A. Simulation parameters

An extensive simulation has been carried out on the proposed scheme to evaluate its performance in different network conditions and topologies. The proposed scheme is implemented on network simulator *ns-2*. The simulation was carried out on an abstracted MANET that consisted of 100 mobile nodes distributed over an area of 1500 m x 1500 m. The duration of simulation was 120 s. Random way-point model has been chosen for node mobility pattern with maximum speed of a node as 10 m/s and average host pause time of 30 s. During each simulation, five communication sessions were established that require certificate exchanges among the nodes. The *dynamic source routing* (DSR) protocol was used as the routing protocol [21]. For each configuration, 10 simulations runs were executed and the computed average value is taken as the final observation. The following three metrics are studied:

(1) *Valid public key acceptance rate*: it is the ratio of the number of valid public keys accepted and the total number of public keys requested for by the nodes in the MANET.

(2) *Corrupted public key acceptance rate*: it is the ratio of the number of corrupted public keys accepted and the total number of public keys requested for by the nodes in the MANET.

(3) *Delay*: it is the interval between the time a request is sent for a public key and the time when the public key is accepted as valid.

In the simulation, attacks are simulated where the attacker nodes send spurious certificates to the nodes which have requested for those certificates. These attacks can be isolated attacks where every attacker certifies a different public key. However, the attackers may also launch a cooperative attack where a group of attackers collude and send certifications for the same public key that is spurious. Both these types of attacks- isolated and collusion- are simulated. The percentage of attacker nodes is varied from 0% to 40% of the total number of nodes in the network. Node initialization at the network bootstrapping phase is not simulated. It is assumed that each node has successfully executed the initialization step by exchanging requisite number of certificates with the honest nodes in the network. The number of certificate exchanged during the initialization is varied from 0 to 20 for each source and destination. A trust value of 0.75 is assigned to a node that is authenticated during the initialization step, while other nodes are assumed to have a trust value of 0.5. MPKTV is varied from 0.5 to 0.9.

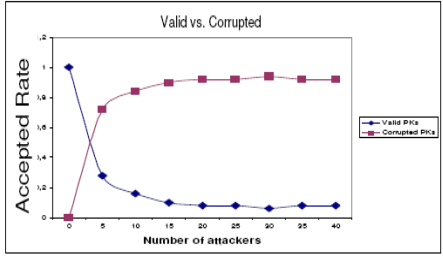

Figure 3. Public key acceptance rate for varying number of attackers

The number of certificate exchange during the initialization varies from 0 to 20 for each source-destination pair. A trust value of 0.75 is assigned to any node authenticated during the initialization, while a value of 0.5 is assigned to other nodes. The MPKTV varies form 0.5 to 0.9.

### B. Analysis

1. *Isolated attackers*: Figure 3 depicts the variation of the valid public key acceptance rate and the corrupted public key acceptance rate with varying number of attackers. The MPKTV is kept constant at 0.5 while the number of attackers is varied from 0 to 40. There was no initial trust between any pair of nodes in the network. It is observed that the rate of valid public key acceptance falls rapidly as the number of attacking nodes increases. The trend is just the reverse for the rate of corrupted public key acceptance rate. Since there was no initial trust at the initial stage, no intermediate node could issue a certificate for a requesting node. Only the destination node could reply to a CREQ message. With the increase in number of attacking nodes, there is a higher probability that an attacker sends a reply to a CREQ message. Since MPKTV is taken 0.5, every public key is considered as valid and accepted by the requester. For higher values of MPKTV, no public key is accepted since no node has sufficient trust level for issuing an acceptable certificate.

Figure 4(a), (b) and (c) show the valid public key acceptance rate with 5, 10 and 20 known nodes in the network respectively for different values of MPKTV. It is observed that except when MPKTV = 0.5, the acceptance rate is as high in all cases. It has also been observed that the acceptance rate of valid public key increases by 10% when the number of initially known nodes is increased form 5 to 10 and then from 10 to 20. With more number of nodes initially known, more nodes send replies to a CREQ message, the average trust in a CREP message increases and thus more public keys are accepted. When MPKTV value is 0.5, any reply is accepted, and the probability to receive a valid public key increases. As the number of attackers increases and becomes more than the number of nodes initially known, the probability of accepting corrupted public keys increases.

Figure 5 shows the delay associated in receiving the reply to a CREQ in absence of any attacker node in the network. As the number of known nodes increases, the time required to receive a sufficient number of replies decreases to satisfy a given MPKTV. Moreover, the delay increases with MPKTV because of the increasing requirement for acceptance of a public key. However, a node that has many certifiers of its public key will be quickly authenticated even when the MPKTV is high. This is validated in Figure 5 as delay is found to decrease with increase in number of nodes known initially in the network.

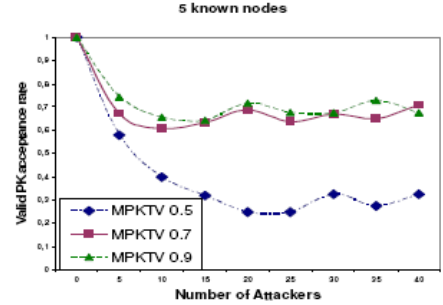

Figure 4(a). The acceptance rate of valid public key with 5 known nodes

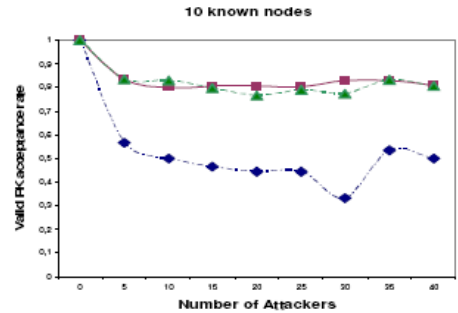

Figure 4(b). The acceptance rate of valid public key with 10 known nodes

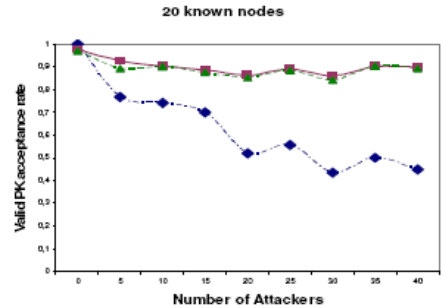

Figure 4(c). The acceptance rate of valid public key with 20 known nodes

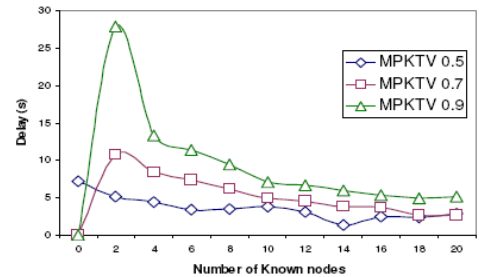

Figure 5. Delay is certification for varying number of known nodes

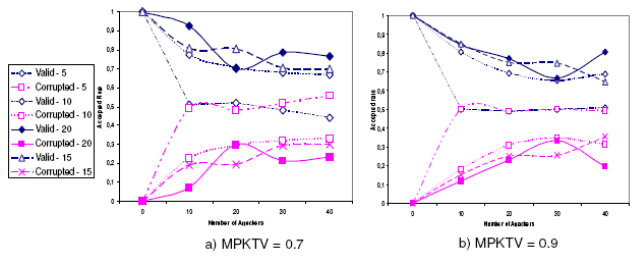

Figure 6. Performance in MANET attacked by a group of colluding nodes (MPKTV= 0.7 in (a) and 0.9 in (b))

*2. Colluding attackers*: Finally, the proposed protocol is simulated in a scenario with colluding attacker nodes. In Figure 6 (a) and Figure 6 (b), *valid-x* and *corrupted-x* denote the rate of acceptance of valid and corrupted certificates respectively when *x* nodes are known initially in the network. It may be observed that some corrupted public keys are accepted since a sufficient number of colluding attackers issue certificates for these corrupted public keys. With MPKTV = 0.9 the rate of acceptance of corrupted public keys is less but as expected, it increases with the number of attackers. Similarly, the rate of acceptance of valid certificates increases with the increase in the known nodes in the network. Nevertheless the increases are much more important when the number of known nodes goes from 5 to 10 that for other increases. An interesting point to note is that with 20 nodes known to the source and the destination, the acceptance rate of corrupted certificates decreases even when the number of attackers increases from 30 to 40. This is because of the fact that when the network contains many attackers, source and destination are more likely to have known nodes in common. Since the common nodes are safe certifiers, with more common nodes in the network, higher is the probability of safe certificate exchange.

## V. CONCLUSION

In this paper, a key exchange protocol for MANETs is proposed that can be integrated with a routing protocol. The protocol is light-weight, efficient and alleviates the routing-security interdependency cycle. Simulation results show that the protocol is resistant to isolated attack launched by malicious nodes that may introduce spurious certificates in the networks. It also performs well against cooperative attacks when sufficient level of trust exists among some nodes before the network deployment.